\documentclass[aps,prl,twocolumn,groupedaddress]{revtex4}
\usepackage{graphicx}
\begin{document}
\title{Square vortex lattice at anomalously low magnetic fields\\ in electron-doped Nd$_{1.85}$Ce$_{0.15}$CuO$_{4}$}
\author{
R. Gilardi,$^1$ J. Mesot,$^1$ S. P. Brown,$^2$ E. M. Forgan,$^2$ A.
Drew,$^3$ S. L. Lee,$^3$ 
R. Cubitt,$^4$  C. D. Dewhurst,$^4$ T. Uefuji,$^5$  K. Yamada$^5$
}
\affiliation{
$^1$ Laboratory for Neutron Scattering, ETH Zurich and PSI Villigen, CH-5232 Villigen PSI, Switzerland\\
$^2$ School of Physics and Astronomy, University of Birmingham, Birmingham B15 2TT, UK\\
$^3$ School of Physics and Astronomy, University of St. Andrews, Fife, KY16 9SS, UK\\
$^{4}$ Institut Laue-Langevin, BP 156, F-38042 Grenoble, France\\
$^{5}$ Institute for Materials Research,Tohoku University, Sendai 980-8577, Japan }
\begin{abstract}
We report here on the first direct observations of the vortex
lattice in the bulk of electron-doped
Nd$_{1.85}$Ce$_{0.15}$CuO$_{4}$ single crystals. Using small angle
neutron scattering, we have observed a square vortex lattice with the
nearest-neighbors oriented at 45$^{\circ}$ from the Cu-O bond
direction, which is consistent with theories based on the $d$-wave
superconducting gap. However, the square symmetry persists down to
unusually low magnetic fields. Moreover, the diffracted intensity
from the vortex lattice is found to decrease rapidly with
increasing magnetic field. 
\typeout{polish abstract}
\end{abstract}
%\date{\today}
%\pacs{74.25.Qt, 61.12.Ex, 74.72.Jt, 74.20.Rp}
%
\maketitle
\begin{figure}[t]
\hspace{-1mm}
\includegraphics[width=6.8cm]{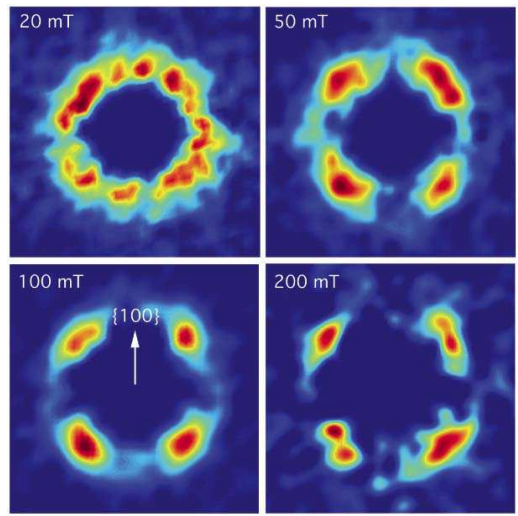}
\caption{SANS diffraction patterns measured at $T
\approx 2.5$~K, after field cooling from 30~K in $B=$~20,
50, 100 and 200~mT. A background taken at
$T=30$~K has been subtracted. The \{100\} directions, corresponding to
the Cu-O bond directions in the CuO$_2$ planes, are oriented vertical/horizontal in the pictures.
}
\label{fig1}
\end{figure}
\begin{figure}[t]
\includegraphics[width=6.8cm]{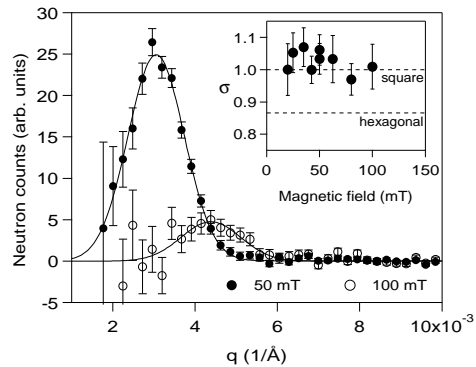}
\caption{Tangential sum of the neutron signal vs wavevector $q$ for $B=$~~50
and 100~mT at $T\approx 2.5$~K. Inset: field dependence of the coordination-dependent quantity
$\sigma$ (see Eq.\ref{sigma}) at low fields; the horizontal lines indicate the expected values for
hexagonal and square VL. }
\label{fig2}
\end{figure}

%\narrowtext
%
It is a matter for debate whether hole-doped and electron-doped
high-$T_c$ cuprate superconductors (HTSC) can be described within
a unified physical picture~\cite{Yeh,Manske}. Indeed,
electron-doped HTSC have markedly different properties from
hole-doped HTSC. For example, electron-doped materials have
comparatively low values of the superconducting transition
temperature, $T_c $, and much lower values of the upper critical field,
$B_{c2}$. Furthermore, their normal-state resistivity varies as
$T^2$ as expected for a Fermi-liquid~\cite{Hidaka,Tsuei1}, and the
presence of a pseudogap is still under discussion~\cite{Alff1}.
Electron-doped HTSC also appear much closer to long-range
antiferromagnetic (AF) order, which can in fact coexist with
superconductivity ~\cite{Kang,Yamada,Fujita,Uefuji}. In hole-doped HTSC,
the $d$-wave nature of the order parameter is well-established
\cite{Tsuei2}. However, the evidence for the symmetry of the
superconducting gap in electron doped materials (which has
important implications for the pairing mechanism~\cite{Manske}) is
somewhat contradictory. Earlier measurements of the penetration
depth~\cite{Wu} and tunneling experiments
\cite{Kashiwaya} supported {\it s}-wave symmetry, whereas
a {\it d}-wave superconducting order parameter is indicated by more recent phase-sensitive~\cite{Tsuei} and ARPES
experiments~\cite{Sato,Armitage}. The electron-doped
superconductors are of particular interest in this respect, since
they have a tetragonal structure (rather than orthorhombic) and
therefore should show pure {\it d}-wave behavior, unaffected by
admixture of an {\it s}-wave component associated
with the orthorhombicity~\cite{Tsuei2}. \newline\indent
Recently, there
has been considerable interest in the nature of the vortex lattice
(VL) in unconventional superconductors.
For instance, vortex cores in $d$-wave superconductors are predicted to have a distinctive
four-fold structure~\cite{Berlinsky,Franz,Affleck,Shiraishi,Ichioka}. 
This leads to the expectation that a square VL is formed at high magnetic field,
with the nearest-neighbor directions aligned with the nodes of the order
parameter.
These theoretical predictions are consistent with the small-angle
neutron scattering (SANS) observation of a transition from an
Abrikosov-like hexagonal VL to a square VL in the
hole-doped HTSC YBa$_2$Cu$_3$O$_7$ (YBCO)~\cite{Brown}. However,
similar measurements on overdoped
La$_{2-x}$Sr$_x$CuO$_4$ (LSCO)~\cite{Gilardi} show a square
lattice with nearest neighbors oriented at 45$^{\circ}$ to the
nodes of the superconducting order parameter. The orientation of the nodal directions in the heavy-fermion
superconductor CeCoIn$_5$ is at present uncertain \cite{Aoki}, so it is unclear if the recent SANS observations of the VL in this material \cite{Eskildsen} confirm theoretical expectations.
It is therefore of great interest to gain further information from another class of superconductors, such as the electron-doped HTSC. Moreover, these compounds do not have twin
planes present in orthorhombic systems such as YBCO and LSCO. These planar defects with a suppressed order parameter are capable of pinning the VL \cite{Yethiraj}.
Finally, the low values of the upper
critical field ($B_{c2}\sim$ 10~T compared to $B_{c2}\sim$ 100~T
in hole-doped HTSC) allow the investigation of the whole magnetic
phase diagram.\newline\indent
Only recently could large enough electron-doped crystals be
produced, and to our knowledge, a SANS investigation of their VL
has not yet been published. We report here the first direct
observation of a VL in Nd$_{1.85}$Ce$_{0.15}$CuO$_{4}$ (NCCO). Our
SANS experiments (see ref.~\cite{Forgan} for a description of the technique) were performed on the instrument D22 at the Institut Laue Langevin, France, using neutrons with a wavelength
$\lambda_n$= 6 \AA - 20 \AA. Crystals of NCCO were grown in a
mirror furnace and annealed as described in ref.~\cite{Kurahashi}
to give an onset $T_c \approx$ 25 K ($\Delta T_c \approx$ 3 K,
10\%-90\% criterion). Samples with two shapes were investigated in applied
magnetic fields up to 0.4 T. In high fields, a cylinder of 5 mm
diameter, consisting of two nearly aligned crystals, was
mounted in a cryostat with the magnetic field direction bisecting
the two ${\bf c}$-directions and at 7$^{\circ}$ to each of them.
Single crystal plates of $\sim$1.5 mm thickness were used at low fields: this required a long neutron
wavelength and the smaller thickness reduced the effects of
neutron absorption in the sample. They were mounted so that the ${\bf
c}$-direction was within $\sim 2 ^{\circ}$ of the applied field,
which was approximately parallel to the incident beam. There were
no significant differences in the VL diffraction patterns obtained
from different samples in measurements at the same field. 
\newline\indent In Fig.~\ref{fig1} we show the VL diffraction patterns
obtained at various magnetic fields. A background measured above $T_c$ has been subtracted because the VL signal is extremely weak ($\sim$0.013\% of the total detector counts at 50 mT). For magnetic fields larger than
50~mT, the VL clearly has square coordination. The
nearest-neighbor directions are parallel to the \{110\} crystal
directions, which correspond to the nodes of the superconducting
order parameter, in agreement with predictions for {\it d}-wave
HTSC \cite{Berlinsky,Franz,Affleck,Shiraishi,Ichioka}. Further
evidence for a square VL is given by the positions of the Bragg
spots in reciprocal space. For first-order diffraction, the relationship between the magnitude $q$ of the wavevector and the magnetic field $B$ is given by
flux quantization:
\begin{equation}
q=2\pi\sqrt{\frac{B}{\sigma\Phi_0}},
\label{sigma}
\end{equation}
where $\sigma$ is equal to 1 or $\sqrt{3}/2$ for square and
hexagonal lattices respectively, and $\Phi_0$ is the flux quantum. $q$ can be
obtained by fitting the tangential sum of the neutron signal
with a Gaussian, as shown in Fig.~\ref{fig2}. As expected, the
position of the peak shifts to higher $q$ with increasing magnetic
field (confirming the VL origin of the neutron signal), and the
extracted values of $\sigma$ are consistent with a square VL at \textit{all} fields measured (see
inset to Fig.~\ref{fig2} for the lower fields). We have confirmed by magnetization measurements
that the value of the trapped flux at low fields is within $\sim$~1\% of the applied field. Hence, flux expulsion has an insignificant effect on the value of the $q$-vector.
\begin{figure}[t]
\hspace{-1mm}
\includegraphics[width=6.8cm]{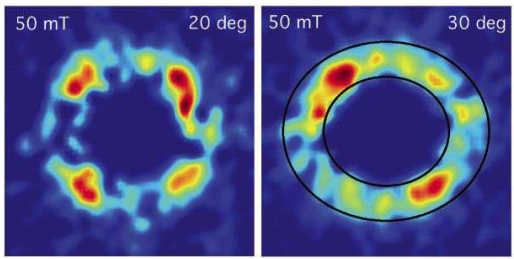}
\caption{SANS diffraction patterns at $B=$~50~mT taken as in Fig.~\ref{fig1} but with the ${\bf c}$-direction rotated about the vertical axis by 20$^{\circ}$ and 30$^{\circ}$ to the field direction. The growth direction and long axis of the sample lay 15$^{\circ}$ counterclockwise to the vertical and may be associated with the stronger intensity in these quadrants. The ellipses drawn have axial ratio of cos(30$^{\circ}$).
}
\label{fig3}
\end{figure}
\begin{figure}[t]
\includegraphics[width=6.8cm]{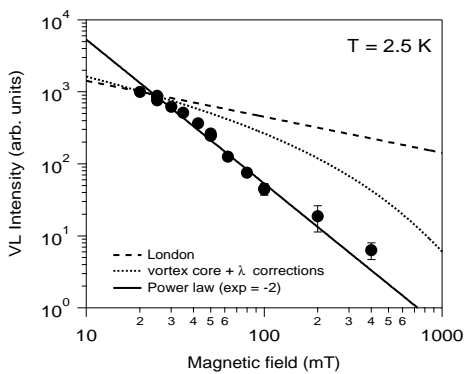}
\caption{VL intensity measured at $T\approx$ 2.5 K as a function of
magnetic field. The dotted line is the expected field dependence
taking into account core effects and 
field-dependence of the penetration depth (see text).}
\label{fig4}
\end{figure}
\newline\indent As shown in Fig.~\ref{fig1}, the VL orientation is not perfect at any field.
However, the orientation becomes rather more disordered as the field
is decreased
below 50~mT and at 20~mT the intensity distribution becomes ring-like.
A similar distribution of intensity at low magnetic fields has been
observed in LSCO
\cite{Gilardi}, and was attributed to the superposition of
diffraction patterns
from various domain orientations of {\it hexagonal} coordination,
since the value of
$\sigma$ at low fields was consistent with that of an hexagonal VL.
Moreover, by rotating the ${\bf c}$-axis 10$^{\circ}$ away from the
field direction,
the degeneracy of the VL system could be reduced, and the hexagonal
coordination
of the VL in LSCO was confirmed.
In NCCO, on the contrary, the values of $\sigma$ at low magnetic
fields are still
consistent with that of a square VL, and
measurements at 20 mT
with the ${\bf c}$-axis rotated 
 away from the field direction 
caused little change in the pattern.
The effects of rotating of the field away from the ${\bf c}$-axis at 50 mT are shown in Fig.~\ref{fig3}.
At 20$^{\circ}$ little change is observed, whereas at 30$^{\circ}$ the pattern becomes rather disordered and the scattered intensity lies on an ellipse as expected from the uniaxial anisotropy of the crystal. However the average value of $\sigma$ is unchanged. It is surprising that the vortex lattice maintains square coordination without being aligned to the crystal lattice.
\newline\indent In addition to the $d$-wave scenario for square VL coordination, one should also consider  the effects of Fermi surface anisotropy. An appropriate theory for large-$\kappa$ materials well below $B_{c2}$ is London theory with non-local corrections \cite{Kogan}, which has been extensively used to account for VL phase transitions in the borocarbides \cite{Eskildsen2,Levett}.
If both these effects are present \cite{Nakai}, we would expect the strong angular variation of the $d$-wave gap to dominate over the usually smaller variation of the Fermi velocity. 
ARPES experiments together with band structure
calculations~\cite{Damascelli} indicate that  NCCO and YBCO  both have
nearly isotropic hole-like Fermi surfaces with a slight four-fold distortion oriented so as to favor the observed square VL orientation. In overdoped LSCO, on the other hand, the
Fermi surface is electron-like~\cite{Yoshida,Ino} with a square shape oriented at
45$^\circ$ to that of the other two compounds.
Moreover LSCO exhibits a pronounced anisotropy in the Fermi velocity. Both the shape of the Fermi surface and the Fermi velocity anisotropy would indicate (via nonlocal
effects) the VL orientation actually observed in LSCO~\cite{Gilardi}. 
\newline\indent However, both $d$-wave and nonlocal effects should only be important at fields which are substantial fraction of $B_{c2}$, when the inter-vortex spacing is comparable to the coherence length. 
For instance, $d$-wave calculations indicate that the square symmetry has lower free energy than the hexagonal one at applied fields greater than 0.15\thinspace$B_{c2}\approx$
1.2~T ~\cite{Ichioka}, or than $B_{c2}/\kappa\approx$
0.4~T~\cite{Shiraishi} (taking the values of $B_{c2}\approx$
8~T~\cite{Herrmann}, penetration depth $\lambda_L \approx$ 1250 \AA
~\cite{Anlage} and coherence length $\xi\approx$ 60
\AA~\cite{Herrmann}, therefore $\kappa=\lambda_L/\xi \approx$ 20).
Although the characteristic fields in electron-doped HTSC are generally lower than in the hole-doped materials, these estimates are too large to explain our results.  
Hence, our observation of a square VL in NCCO down to very low magnetic fields is 
rather surprising, unless another source of anisotropy in the CuO$_2$ planes is present.
One candidate for this is the Cu antiferromagnetic correlations, whose characteristic wavevector \cite{Yamada} coincides in direction with the VL reciprocal lattice wavevector. 
\newline\indent Another matter of great
interest is the variation of the diffracted intensity with field.
The intensity $I_{hk}$ of a single $(h,k)$
reflection (integrated over the rocking curve of the VL) is given
by \cite{Christen}:
\begin{equation}
I_{hk}=2\pi\phi\Big(\frac{\mu}{4}\Big)^2\frac{V\lambda_n^2}{\Phi_0^2q_{hk}}|F_{hk}|^2
\propto \frac{|F_{hk}|^2}{q_{hk}}, \label{eq1}
\end{equation}
where $\phi$ is the incident neutron flux, $\mu$ is the neutron
magnetic moment,
$V$ is the sample volume and $q_{hk}$ is the $(h,k)$ reciprocal lattice vector.
$F_{hk}$ is the ``form factor'' of the $(h,k)$ reflection. It is a Fourier
component of the spatial variation of the magnetic field, and in the
London limit
is related to the penetration length $\lambda_L$ by:
\begin{equation}
F_{hk}=\frac{B}{1+(q_{hk}\lambda_L)^2}
\label{eq2}
\end{equation}
For $B_{c1} \ll B \ll B_{c2}$, the second term in the denominator
is dominant and this gives $I_{hk} \propto q_{hk}^{-1}
\propto B^{-\frac{1}{2}}$. In our NCCO samples, the rocking curves
of the VL diffracted intensity were found to be flat, over a range of $\pm 4^{\circ}$, within experimental error. 
The expected width of the rocking curves can be estimated using Eqns. (\ref{eq1}) and (\ref{eq2}) and the value of $\lambda_L$ quoted in $\mu$SR measurements \cite{Anlage}. At 50 mT we obtain a width larger than 10$^{\circ}$, consistent with our experimental observations.
We therefore
measured the intensity as a function of field at a fixed sample
angle. Assuming that the rocking curve width remains constant with
field, the measured intensity is proportional to the integrated
intensity. In contrast with the London prediction, we observe a
strong field-dependence of the scattered intensity (see
Fig.~\ref{fig2} and Fig.~\ref{fig4}), which becomes immeasurably small above $B\approx$
0.4~T, which is well below $B_{c2}$. If represented by a power
law, this variation has an exponent of about~-2.
(If the rocking curve becomes narrower with increasing field, then the field dependence of the VL intensity becomes even more rapid and more difficult to interpret. A broadening of the rocking curve as a function of increasing field would be consistent with increasing disorder as discussed below).\newline\indent
We consider two possible intrinsic reasons for the fast decrease
of the VL intensity with field.  One is the effect of the finite
size of the vortex core, which will lead to deviations from the
London predictions at large values of the wavevector $q$. This
effect can be modeled by a calculation using GL
theory~\cite{Brandt}, giving an algebraic expression for the
correction to the form factor: $ exp(-\sqrt{2}q\xi)$ which may
also be written as
 $exp(-2(\pi B/B_{c2})^{\frac{1}{2}})$. In addition,
there is expected to be a specifically $d$-wave effect -
an increase of the penetration depth with field~\cite{Amin}.
$\mu$SR measurements in YBCO
indicate that the variation of $\lambda$ with $B$ is linear at low
fields ($B <$ 2~T)~\cite{Sonier2}:
\begin{equation}
\frac{\lambda(B)}{\lambda(0)}=1+\beta\frac{B}{B_0},
\end{equation}
where $\beta$ is a temperature-dependent coefficient that remains
finite at $T$=0 K,
and $B_0$ a characteristic field on the order of the thermodynamic critical
field $B_c$. 
If we assume that a similar linear dependence of $\lambda(B)$ is
valid for NCCO, and
include the core-size effects as well, we can model the
variation of the VL intensity with field:
\begin{equation}
I_{hk} \sim B^{-\frac{1}{2}}\Big(1+\beta\frac{B}{B_0}\Big)^{-4}
exp(-4 (\pi B/B_{c2})^{\frac{1}{2}})
\end{equation}
\indent For NCCO we used $\beta =  7 \cdot 10^{-2}$ (as for YBCO) and
$B_0 \sim B_{c2}/\sqrt{2}\kappa = 0.28$~T. As can be seen in Fig.~\ref{fig4},
we cannot represent the strong field-dependence of our experimental
data by these intrinsic effects.
It may be that an enhancement of AF correlations by the applied field \cite{Kang} is reducing the superfluid density. However, we find that even doubling $\beta$ and the exponential prefactor does not reproduce our results. 
The rapid decrease of the VL intensity is more likely to be due
to a transition to a more disordered vortex system, as has been predicted
theoretically~\cite{Giamarchi} and
associated with a second peak in magnetization
measurements ~\cite{Giller}.
SANS experiments in the isotropic (K,Ba)BiO$_3$ system
\cite{Joumard} revealed a similar rapid loss of diffracted intensity near the second peak. 
In hole-doped Bi$_{2.15}$Sr$_{1.95}$CaCu$_2$O$_{8+x}$, a strong
decrease of the VL intensity with increasing field was attributed
to a crossover from a 3D to a 2D vortex system \cite{Cubitt}. In
NCCO, however, such a dimensional crossover is expected to occur
at much higher magnetic field $B_{2D}$=$\Phi_0/(\gamma s)^2\sim$
13~T ($>B_{c2}$!), where s $\sim$ 6 \AA~is the distance between
CuO$_2$ planes and $\gamma\sim$ 21 \cite{Hidaka} is the
anisotropy. In this respect, NCCO seems to be similar to
underdoped LSCO~\cite{Divakar}, in which $\mu$SR measurements have
given clear evidence of a field-induced crossover to a more
disordered, but still three-dimensional VL.
\newline\indent In conclusion, we have made the first direct observation of the VL
in the electron-doped NCCO, which is the first tetragonal
HTSC to be investigated by SANS. Contrary to theoretical
expectations, the VL remains square down to very small fractions
of $B_{c2}$. In addition, we have observed an unusually fast decrease
of the VL intensity with increasing magnetic field, which is
probably due to a crossover to a more disordered vortex state.
\newline\indent This work was supported by the Swiss National Science Foundation,
the Engineering and Physical Sciences Research Council of the U.K.
and the Ministry of Education and Science of Japan.


\begin{references}

\bibitem{Yeh}
N.-C. Yeh and C.-T. Chen, Int. J. Mod. Phys. B \text{17}, 3575 (2003).

\bibitem{Manske}
D. Manske \textit{et al.}, Phys. Rev. B \textbf{62}, 13922 (2000).

\bibitem{Hidaka}
Y. Hidaka and M. Suzuki, Nature \textbf{338}, 635 (1989).

\bibitem{Tsuei1}
C.C. Tsuei \textit{et al.}, Physica C \textbf{161}, 415 (1989).

\bibitem{Alff1}
L. Alff \textit{et al.}, Nature \textbf{422}, 698 (2003).

\bibitem{Kang}
H.J. Kang \textit{et al.}, Nature \textbf{423}, 522 (2003).

\bibitem{Yamada}
K. Yamada \textit{et al.}, Phys. Rev. Lett. \textbf{90}, 137004 (2003).

\bibitem{Fujita}
M. Fujita \textit{et al.}, Phys. Rev. B \textbf{67}, 014514 (2003).

\bibitem{Uefuji}
T. Uefuji \textit{et al.}, Physica C \textbf{357-360}, 208 (2001).

\bibitem{Tsuei2}
C.C. Tsuei and J.R. Kirtley, Rev. Mod. Phys. \textbf{72}, 969 (2000)
and references therein.

\bibitem{Wu}
D.H. Wu \textit{et al.}, Phys. Rev. Lett. \textbf{70}, 85 (1993).

\bibitem{Kashiwaya}
S. Kashiwaya \textit{et al.}, Phys. Rev. B \textbf{57}, 8680 (1998).

\bibitem{Tsuei}
C.C. Tsuei and J.R. Kirtley, Phys. Rev. Lett. \textbf{85}, 182 (2000).

\bibitem{Sato}
T. Sato \textit{et al.}, Science \textbf{291}, 1517 (2001).

\bibitem{Armitage}
N.P. Armitage \textit{et al.}, Phys. Rev. Lett. \textbf{86}, 1126 (2001).

\bibitem{Berlinsky}
A.J. Berlinsky \textit{et al.}, Phys. Rev. Lett. \textbf{75}, 2200 (1995)

\bibitem{Franz}
M. Franz \textit{et al.}, Phys. Rev. Lett. \textbf{79}, 1555 (1997).

\bibitem{Affleck}
I. Affleck \textit{et al.}, Phys. Rev. B \textbf{55}, R704 (1997).

\bibitem{Ichioka}
M. Ichioka \textit{et al.}, Phys. Rev. B \textbf{59}, 8902 (1999).

\bibitem{Shiraishi}
J. Shiraishi \textit{et al.}, Phys. Rev. B \textbf{59}, 4497 (1999).

\bibitem{Brown}
S.P. Brown \textit{et al.}, Phys. Rev. Lett. \textbf{92}, 067004 (2004).

\bibitem{Gilardi}
R. Gilardi \textit{et al.}, Phys. Rev. Lett. \textbf{88}, 217003 (2002),
Int. J. Mod. Phys. B \textbf{17}, 3411 (2003) and Physica C \textbf{408-410}, 491 (2004).

\bibitem{Aoki}
H. Aoki \textit{et al.}, J. Phys.: Condens. Matter \textbf{16}, L13 (2004) 

\bibitem{Eskildsen}
M.R. Eskildsen \textit{et al.}, Phys. Rev. Lett. \textbf{90}, 187001 (2003).

\bibitem{Yethiraj}
M. Yethiraj \textit{et al.}, Phys. Rev. Lett. \textbf{70}, 857 (1993).

\bibitem{Forgan}
E.M. Forgan in \textit{Neutron Scattering in Layered Copper-Oxide Superconductors}, edited by A. Furrer (Kluwer, 1998) p. 375.

\bibitem{Kurahashi}
K. Kurahashi \textit{et al.}, J. Phys. Soc. Jpn. \textbf{71}, 910 (2002).

\bibitem{Kogan}
V.G. Kogan \textit{et al.}, Phys. Rev. B \textbf{55}, R8693 (1997).

\bibitem{Eskildsen2}
M.R. Eskildsen \textit{et al.}, Phys. Rev. Lett. \textbf{86}, 5148 (2001).

\bibitem{Levett}
S.J. Levett, C.D. Dewhurst and D.McK. Paul, Phys. Rev. B \textbf{66}, 014515 (2002).

\bibitem{Nakai}
N. Nakai \textit{et al.}, Phys. Rev. Lett. \textbf{89}, 237004 (2002).

\bibitem{Damascelli}
A. Damascelli, Z. Hussain and Z.-X. Shen, Rev. Mod. Phys. \textbf{75},
473 (2003) and references therein.

\bibitem{Yoshida}
T. Yoshida \textit{et al.}, Phys. Rev. B \textbf{63}, 220501(R) (2001).

\bibitem{Ino}
A. Ino \textit{et al.}, Phys. Rev. B \textbf{65},
094504 (2002)

\bibitem{Herrmann}
J. Herrmann \textit{et al.}, Phys. Rev. B \textbf{54}, 3610 (1996).

\bibitem{Anlage}
S.M. Anlage \textit{et al.}, Phys. Rev. B \textbf{50}, 523 (1994).

\bibitem{Christen}
D.K. Christen \textit{et al.}, Phys. Rev. B \textbf{15}, 4506 (1977).

\bibitem{Brandt}
A. Yaouanc, P. Dalmas de Reotier and E.H. Brandt,
Phys. Rev. B \textbf{55}, 11107 (1997).

\bibitem{Amin}
M.H.S. Amin, I. Affleck and M. Franz, Phys. Rev. B \textbf{58}, 5848 (1998)
and Phys. Rev. Lett. \textbf{84}, 5864 (2000).

\bibitem{Sonier2}
J.E. Sonier \textit{et al.}, Phys. Rev. B \textbf{55}, 11789 (1997).

\bibitem{Giamarchi}
T. Giamarchi and P. Le Doussal, Phys. Rev. B \textbf{55}, 6577 (1997).

\bibitem{Giller}
D. Giller \textit{et al.}, Phys. Rev. Lett. \textbf{79}, 2542 (1997).

\bibitem{Joumard}
I. Joumard \textit{et al.}, Phys. Rev. Lett. \textbf{82}, 4930 (1999).

\bibitem{Cubitt}
R. Cubitt \textit{et al.}, Nature \textbf{365}, 410 (1993).

\bibitem{Divakar}
U.K. Divakar \textit{et al.}, Phys. Rev. Lett. \textbf{92}, 237004 (2004).

\end{references}
\end{document}